\documentclass[aps,preprint]{revtex4}
\usepackage{graphicx,amssymb}
\begin{document}
\setcounter{page}{1}

\title{ Flux-Pinning Effects and Mechanism  of Water-Quenched  5 wt.\% (Fe, Ti) Particle-Doped MgB$_2$ Superconductor }
\author{H. B. \surname{Lee}}
\email{superpig@pusan.ac.kr}
\thanks{Fax: +82-51-513-7664}
\author{G. C. \surname{Kim}}
\author{Hyoungjeen \surname{Jeen}}
\author{Y. C. \surname{Kim}}

\affiliation{Department of Physics, Pusan National University, Busan 609-735, Korea}

\begin{abstract}
 We have studied magnetic behaviors of water-quenched 5 wt.\% (Fe, Ti) particle-doped MgB$_2$ comparing with those of air-cooled one.  Generally, grain refinement is achieved by water-quenching, 
 which means  increases of grain boundaries in a superconductor.  We inspected that the increased grain boundary density of a superconductor influenced what kinds of flux-pinning effects on the field dependence of magnetization. 
 As a result, grain boundaries are served  as a pinning centers on a higher magnetic field (2.5 - 6.0 Tesla) whereas they are served  as a pathway to facilitate  movements of fluxes pinned on volume defects at a lower field (0.2 - 2.0 Tesla). 
 Inspecting the characteristic of grain boundaries  in superconductor, we explained that they have a flux-pinning effect as well as the flux-penetrating promotion effect. By using TEM, we confirmed that the width of grain boundary in the specimens is approximately 1 nm, which is not wide enough to pin several flux quanta across the width. 
As temperature increases, flux-pinning effect of water-quenched 5 wt.\% (Fe, Ti) particle-doped MgB$_2$ decreases significantly when compared with that of air-cooled one. The behavior was  because flux-pinning effect of grain boundaries decreases and flux-penetrating promotion effect of them increases, which are considered to be caused by increased coherence length of the superconductor.

\end{abstract}

\pacs{74.60.-w; 74.70.Ad}

\keywords{ MgB$_2$, Flux pinning, (Fe,Ti) particles, grain boundary, planar defects, TEM, water-quenching }

\maketitle
\section{Introduction} 
 MgB$_2$ has been known  a superconductor which has a weak field dependence of magnetization \cite{Kimishima, Kunitoshi, Dou}. In order to overcome the weakness, many researchers tried to improve field dependence of magnetization by doping method \cite{Shi, Silva, Chengduo}. We also have been studying the properties of MgB$_2$  doped with (Fe, Ti) particles of which radius is 163 nm on average \cite{Lee2}. 

The best performance was obtained on 5 wt.\% (Fe, Ti) particle-doped MgB$_2$ among the various specimens. Although the specimen showed considerable results as shown in Fig.  \ref{fig2}, we still have a desire that diamagnetic property of the superconductor would increase more on higher field (6.0 Tesla) than that of  air-cooled 5 wt.\% (Fe, Ti) particle-doped MgB$_2$. 
On the other hand, we had a question for what kinds of change occur on  M-H curve if grain boundaries increased under the condition that volume defects are fixed. 
To solve the question, we chose water-quenching (WQing) method.

Generally, if a WQing  is carried out for materials, there is no time for grains to grow due to the rapid cooling rate. Thus,  grains of the material are become finer, and grain boundary (GB) density of the material increased. And increased GBs mean increasing the number of weak pinning sites in the superconductor. The difference of  GBs between specimens causes  field dependence of magnetization curves (M-H curves) to change. 

Up to now, many researchers have studied flux-pinning effects of grain boundaries in superconductors, which are a planar defect 
\cite{Sung, Wang, Dimos, YAMADA, Katase}. 
However, correlations of flux-pinning between grain boundaries and volume defects, and flux-penetrating promotion effects of grain boundaries, were not revealed.
Concerning GBs as  pinning sites, they are not only interconnected, but  also connected on volume defects. 
 Thus,  it is expected that there would be a flux-pinning correlation between volume defects and grain boundaries if pinned fluxes at a volume defects are many.

One noting thing is that a width of GB  and amounts of GBs and  volume defects in MgB$_2$ are different according to fabrication methods. For example, volume defects cannot help increasing when MgB$_2$ was synthesized at lower temperature \cite{Rogado}. 
And widths of GB are different according to fabrication temperature and pressure when the specimen was fabricated in high pressure and high temperature method \cite{Kimishima, Romero}. Therefore, all of data presented in current paper are confined to the MgB$_2$ specimen which was fabricated by nonspecial atmosphere synthesis (NAS) method at 920 $^o$C for 1 hr \cite{Lee}.

In this study,  
 we inspected the mechanism for flux-pinning effects of increased grain boundaries by comparing M-H curves of water quenched 5 wt.\% (Fe, Ti) particle-doped MgB$_2$  with that of air cooled one. 

\section{Experiment}
Pure MgB$_{2}$ and (Fe, Ti) particle-doped MgB$_{2}$ specimens were synthesized using NAS method \cite{Lee}. Briefly,
NAS method needs Mg (99.9\% powder), B (96.6\% amorphous powder), (Fe, Ti) particles and stainless steel tube. 
  Mixed Mg and B stoichiometry, and  (Fe, Ti) particles were added by weight. They were finely ground and pressed into 10 mm diameter pellets. (Fe, Ti) particles were ball-milled for several days, and average radius of (Fe, Ti) particles was approximately 0.163 $\mu$m \cite{Lee4}. 
   On the other hand, an 8 m-long stainless-steel (304) tube was cut into 10 cm pieces. Insert holed Fe plate into the stainless-steel tube. One side of the 10 cm-long tube was forged and welded. The pellets and pelletized excess Mg  were placed at upper layer and lower layer in the stainless-steel tube, respectively. The pellets were annealed at 300 $^o$C  for 1 hour to make them hard before inserting them into the stainless-steel tube. The other side of the stainless-steel tube was also forged. High-purity Ar gas was put into the stainless-steel tube, and which was then welded. After specimens had been synthesized at 920 $^o$C  for 1 hour, they are quenched in water and cooled in air, respectively.  Field dependences of magnetization were measured using a MPMS-7 (Quantum Design). During the measurement, sweeping rates of doped specimens were applied equally for the same flux-penetrating condition. TEM images were taken by 200KV Field Emission Transmission Electron Microscope (TALOS F200X).\\

\section{Results and discussion} 

 \subsection{Effects  on M-H curves by water-quenching of 5 wt.\% (Fe, Ti) particle-doped MgB$_2$ at lower temperatures}
\begin{figure}
\vspace{1cm}
\begin{center}
\includegraphics*[width=12cm]{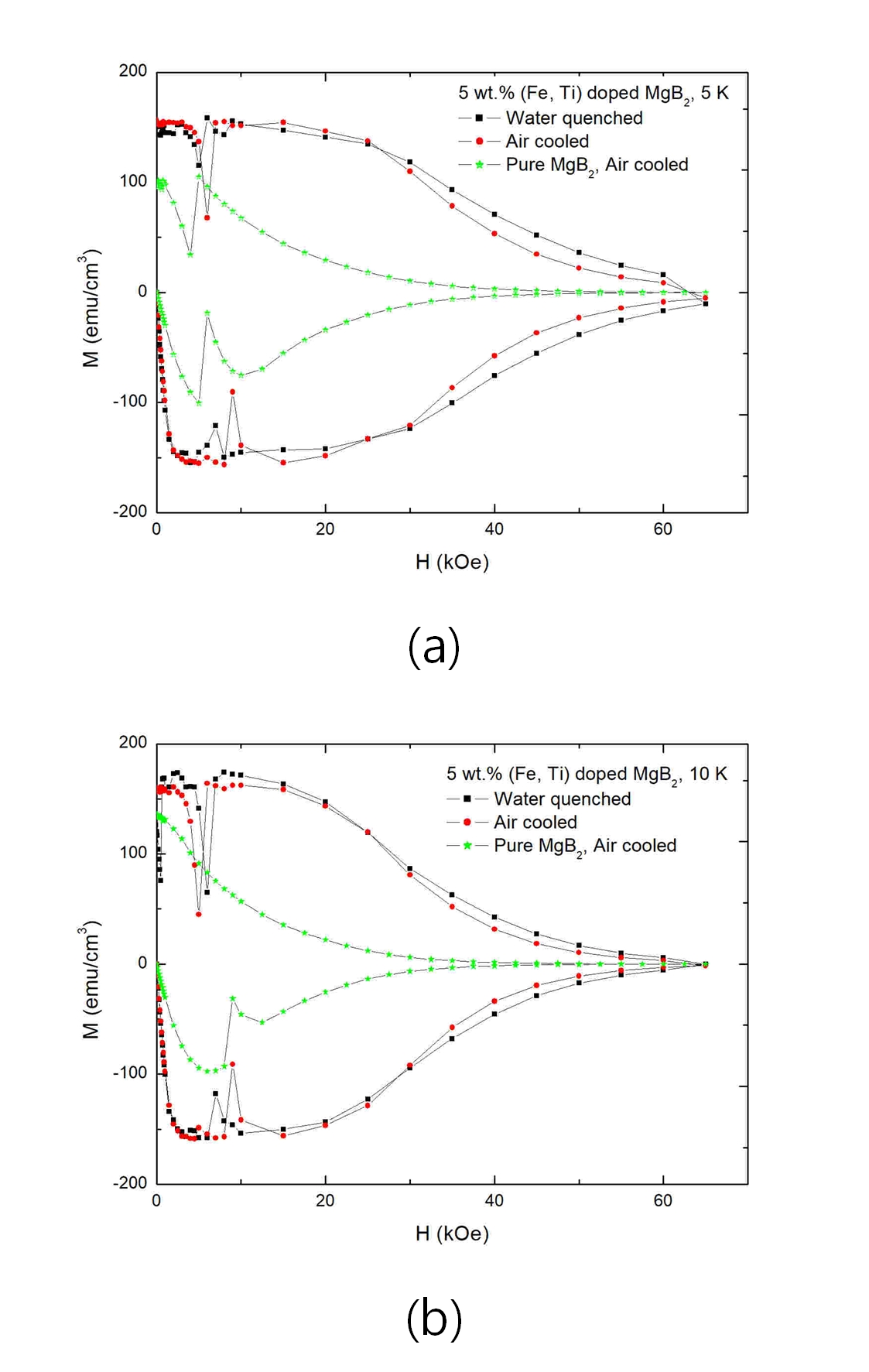}
\end{center}
\caption{ Field dependences of magnetization (M-H curves) of water-quenched  and air-cooled  5 wt.\% (Fe, Ti) particle-doped MgB$_2$ specimens. M-H curve of pure MgB$_2$ is a reference, which was air-cooled. (a): M-H curves at 5 K.  (b): M-H curves at 10 K. } 
\label{fig2}
\end{figure}

Figure \ref{fig2} shows field dependences of magnetization (M-H) curves at 5 K, which are those of 5 wt.\% (Fe, Ti) particle-doped MgB$_2$ specimens cooled in air and quenched in water, respectively. Pure MgB$_2$ are used as a reference, which was cooled in air.  After doping 5 wt.\% (Fe, Ti) particles on MgB$_2$, the field dependences of magnetization of the specimens have been improved greatly regardless of cooling methods. 

On the other hand,  there was a significant difference in the M-H curves between  5 wt.\% (Fe, Ti) particle-doped MgB$_2$ specimens according to cooling methods.
The first is that there was a magnetization difference on the higher applied magnetic field, which is that $\Delta$M of WQed specimen is much larger than the counterpart in magnetic field above 2.5 Tesla (T). 
 And the second is that the crossover field of diamagnetic property of the two specimens was occurred around 2.5 T. It means that the air-cooled (ACed) specimen has a larger $\Delta$M under 2.5 T whereas  WQed specimen has a larger $\Delta$M  over 2.5 T. 

Considering  the reason that $\Delta$M of WQed specimen is smaller in lower field whereas it is larger in higher field compared with that of ACed specimen, this behavior seems to be caused by increased  GBs due to the WQing. As mentioned above, GBs are not only interconnected, but also connected at volume defects. The free energy density of a volume defect is much deeper than that of the planar defect due to its own volume ($\Delta G_{defect}$ = -$\frac{(H-B)^2}{8\pi}\times\frac{4}{3}\pi r^3$, 
where H is applied field, B is magnetic induction and r is radius of a volume defect).

When many fluxes are pinned at a volume defect, the free energy density of the volume defect  would not only increase (depth of $\Delta G_{defect}$ become shallow), but  pinned fluxes at volume defects also have some tension of depinning (pick-out depinning) caused by repulsive force between them \cite{Lee4}. 
Therefore,  they would leak out from the volume defect (leak-out depinning) through the GBs because a volume defects pinned the fluxes many enough and GBs are connected at the  volume defect.  The more GBs are connected at the volume defect, the more some of the pinned fluxes at the volume defect can be leak-out depinned from the volume defect through GBs.

The presence of leak-out depinning fluxes from the volume defect means that the fluxes move into an inside of the superconductor. 
When the behavior occurs, 
magnetic induction (B) increase and diamagnetic property (M) also decreases by 4$\pi$M = B - H. Therefore, diamagnetic property of a WQed  5 wt.\% (Fe, Ti) particle-doped MgB$_2$ having many GBs is lower than that of ACed one on lower field than 2.5 T, as shown in Fig. \ref{fig2}. 

On the other hand, considering the higher applied magnetic field than crossover field ($>$ 2.5 T), all of the volume defects in the superconductor already pinned fluxes to their pinning limits because volume defects pinned the fluxes preferentially, which was caused by the fact that pinned fluxes at volume defects  bent like bow \cite{Lee4}. If applied magnetic field increases more, the volume defect does not pin more fluxes. From this field, GB begins to show its ability as a barrier for the movement of fluxes. Of course, there is also a flux-pinning effect of GBs at lower field than 2.5 T. However, leak-out depinned fluxes from the volume defect through GBs are only visible in  M-H curve  because flux-pinning effect of volume defects is very strong under lower applied magnetic field (especially 0.2 T - 2.0 T in WQed specimen).

The mechanism appeared clearly in the experiments, as shown Fig. \ref{fig2} (a).
 Since the volume defect of ACed specimen begin to be weakened as a pinning center from 1.5 T (2.0 T for WQed specimen) and the pinning effects of GBs are clearly seen over 2.5 T. Considered at 6 T in the figure, it is noted that total pinning effects by GBs of WQed specimen  are more than twice when compared with that of ACed one.

When the external field is less than 1.5 T, M-H curve of ACed 5 wt.\% (Fe, Ti) particle-doped MgB$_2$ is almost parallel with horizontal line, which means $\Delta$H=$\Delta$B region \cite{Lee4},  whereas the  WQed specimen is slightly angled with the horizontal line as shown in Fig. \ref{fig2} (a).   
This behavior of WQed specimen means that the pinned fluxes on the volume defects in the WQed specimen are leak-out depinned from the volume defect more easier than that of the AC specimens owing to increased GBs, and the leak-out depinned  fluxes  penetrate into an inside of the superconductor along GBs. 
 
Therefore, the amount of penetrated fluxes (B) into the superconductor increases, and the diamagnetic property decreases. 
The behavior is profounder at 10 K.
Figure \ref{fig2} (b) shows M-H curves at 10 K, which are that of 5 wt.\% (Fe, Ti) particle-doped MgB$_2$ specimens cooled in air and quenched in water, respectively. The crossover field of the two specimens is occurred around 2.5 T, which is similar with that of 5 K. 

However, inspecting entire M-H curve, the crossover field  shifted a quite higher field. This means that  the influence of flux-pinning effects by GBs on the overall flux-pinning is reduced at 10 K compared with that at 5 K, which also means that pinned fluxes at volume defect leak-out depinned easier at 10 K than 5 K.  The behavior are supported by the results that the diamagnetic  property of 10 K decreased severely  than that of 5 K in WQed doped specimen below crossover field as applied magnetic field increases,  as shown in Fig. \ref{fig2} (a) and (b). 
Therefore, it is determined that the increased GBs by WQing have promoted leak-out depinning as temperature increases.

\subsection{Effects  on M-H curves by water-quenching of 5 wt.\% (Fe, Ti) particle-doped MgB$_2$ at higher temperatures}

\begin{figure}
\vspace{1cm}
\begin{center}
\includegraphics*[width=12cm]{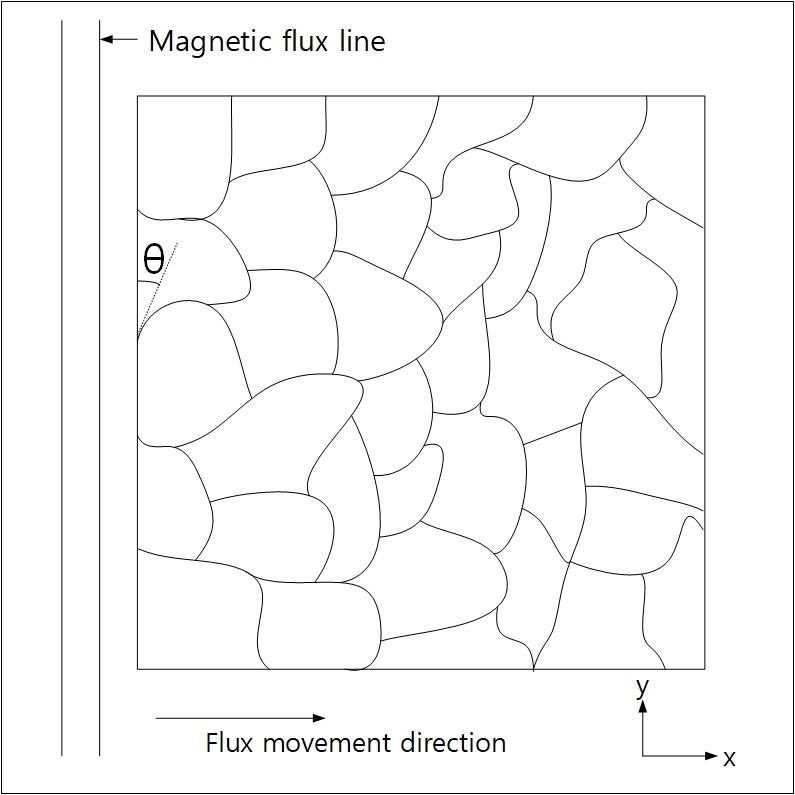}
\end{center}
\caption{Schematic representation of grain boundaries, and different flux-pinning effects of grain boundaries according to thier directions.} 
\label{fig3}
\end{figure}

For proving  flux-pinning effect and flux-penetrating promotion effect of GBs, we assumed that a superconductor has GBs in two dimensions (2-D)  as shown in Fig. \ref{fig3}. When the external magnetic field is applied along y-axis, quantum fluxes in the superconductor move along x-axis  as shown in the figure. They would serve as a pinning center when the fluxes move perpendicular to the GB (when the longitudinal direction of the flux and the plane direction of the GB are parallel). %

  However,  GBs would act as a pathway to penetrate the fluxes into  an inside of  the superconductor rather than flux-pinning sites to pin the fluxes if the fluxes move through GBs that are in parallel with their direction of movement. Thus,  $F_{pinning,GB}$ (pinning force of a GB) in 2-D is  
$F_0 d cos\theta$ if the angle between the flux line and the GB is $\theta$ as shown in Fig. \ref{fig3}, where $F_0$ is the pinning force density of a GB in 2-D when the $\theta$ is 0  
 and $d$ is a width of the GB. On the other hand, $P_{pene,GB}$ (flux-penetrating promotion rate of GB) in 2-D is  $P_{0} d sin\theta$, where $P_0$ is flux-penetrating promotion rate  which is 1 in normal state. 

It is known that coherence length ($\xi$) of a superconductor along temperature follows the equation
\begin{eqnarray}
\xi(T)^2 
\propto\frac{1}{1-t}
 \end{eqnarray} 
 where t = T/T$_c$ \cite{Tinkham}, T$_c$ is critical temperature and T is a temperature which is lower than T$_c$.
Table 1 shows  variations of coherence length along temperature and  ratios of a pinned quantum flux at GB when the width of GB can pin single flux quantum acoss GB. 
It was assumed that H$_{c2}$ of the superconductor is 65.4 T at 0 K and T$_c$ is 37.5 K \cite{Lee6}. 
As shown in the table, flux-pinning effect of GB would drop sharply as  temperature increases. On the other hand, flux-penetrating promotion effect increases as  temperature increases because superconductivity of the specimen decreases and coherence length increases as temperature increases.
Therefore, the movements of pinned fluxes at GB would be much easier as temperature increases. 

\begin{figure}
\vspace{1cm}
\begin{center}
\includegraphics*[width=12cm]{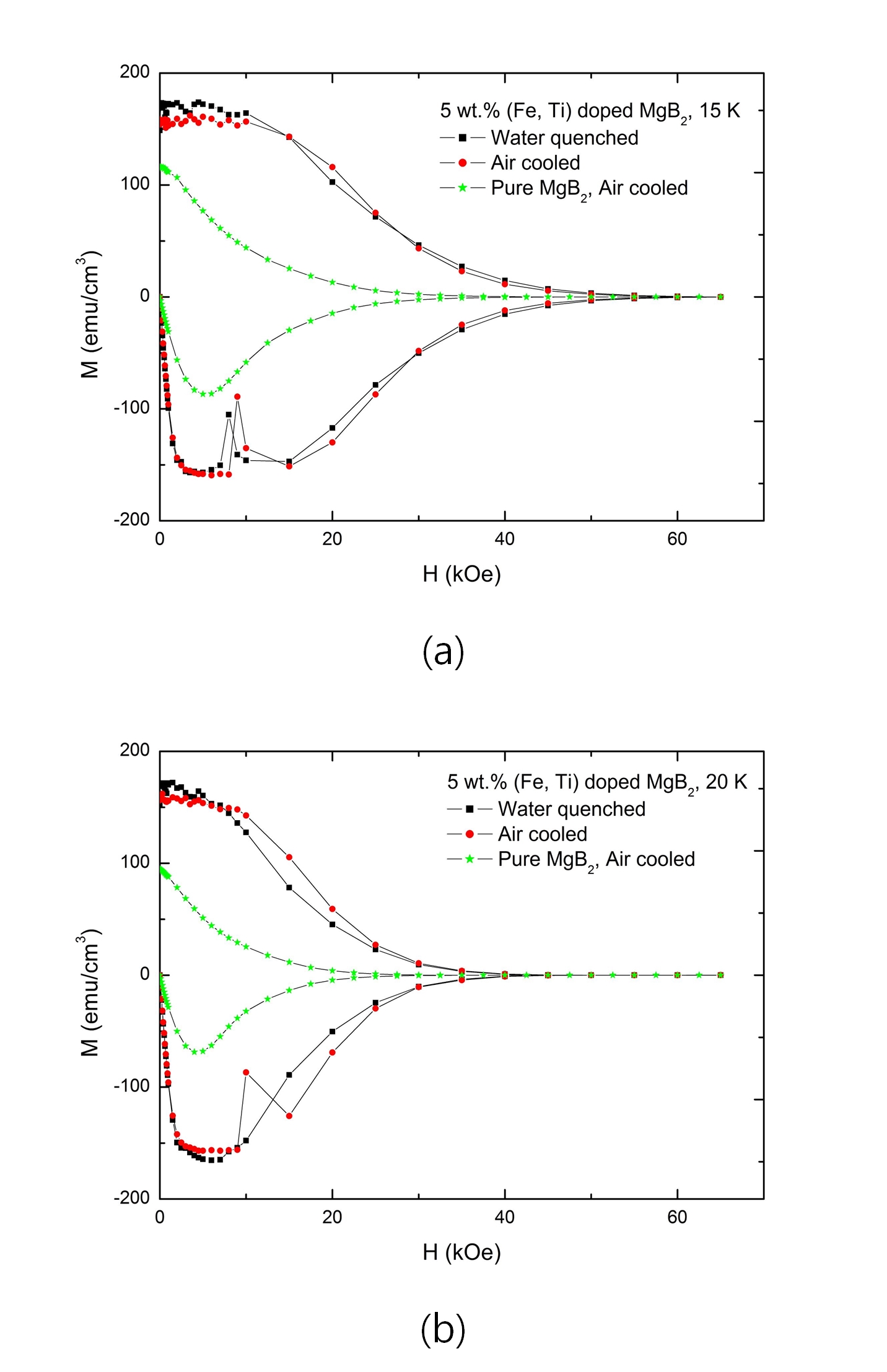}
\end{center}
\caption{Field dependences of magnetization (M-H curves) of water-quenched  and air-cooled  5 wt.\% (Fe, Ti) particle-doped MgB$_2$ specimens. M-H curve of pure MgB$_2$ is a reference, which was air-cooled. (a): M-H curves at 15 K.  (b): M-H curves at 20 K.} 
\label{fig4}
\end{figure}

\begin{figure}
\vspace{1cm}
\begin{center}
\includegraphics*[width=12cm]{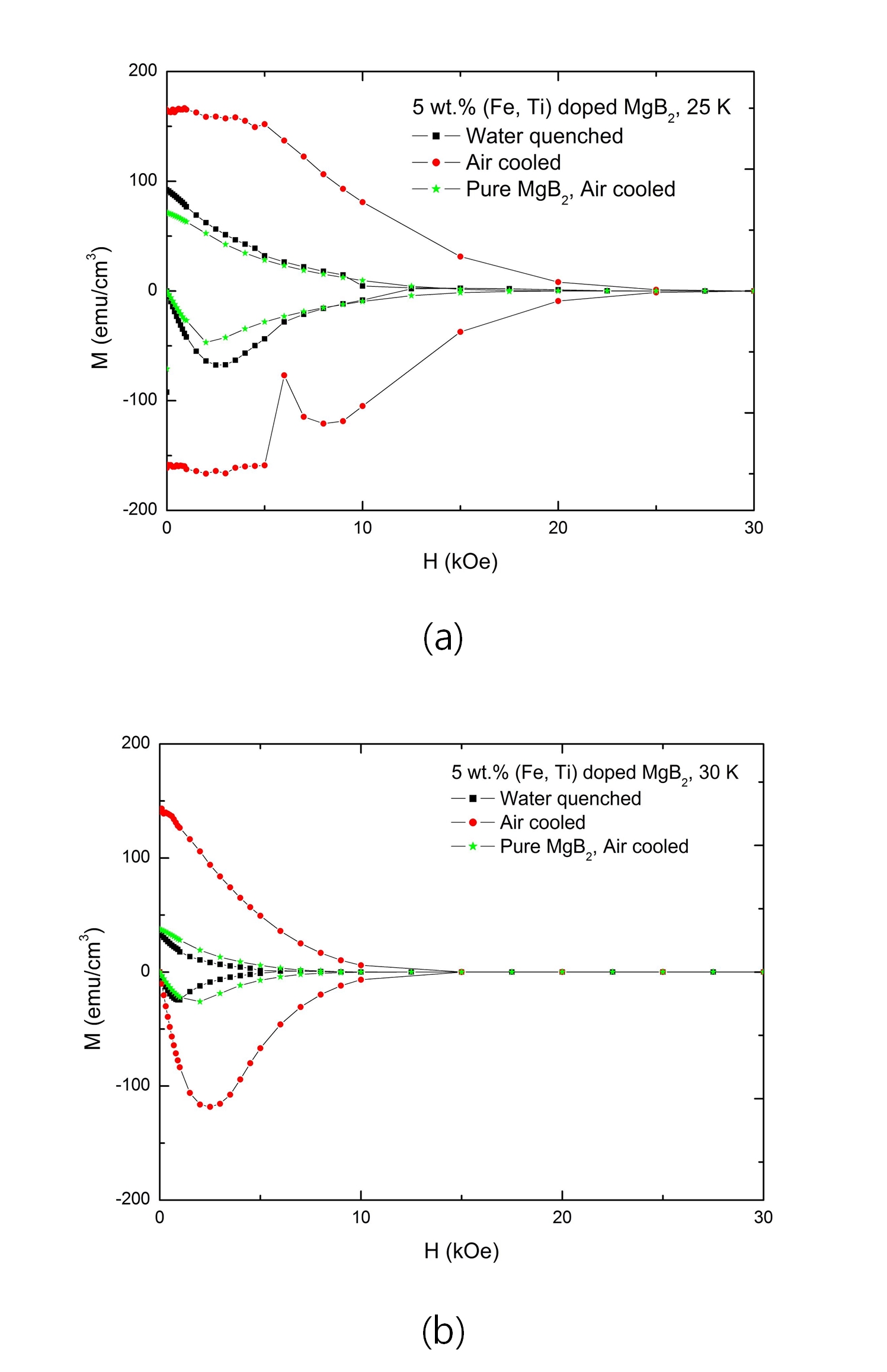}
\end{center}
\caption{Field dependences of magnetization (M-H curves) of water-quenched  and air-cooled  5 wt.\% (Fe, Ti) particle-doped MgB$_2$ specimens. M-H curve of pure MgB$_2$ is a reference, which was air-cooled. (a): M-H curves at 25 K.  (b): M-H curves at 30 K.} 
\label{fig5}
\end{figure}

The behavior predicted by the representation is confirmed by  Fig. \ref{fig4} and Fig. \ref{fig5}.  Figure \ref{fig4} (a) is M-H curves at 15 K, which are those of 5 wt.\% (Fe, Ti) doped MgB$_2$ specimens cooled in air and quenched in water, respectively.  M-H curves of the two specimens almost overlap at at 15 K, but $\Delta$M of the WQed specimen are slightly larger  in the field over 2.5 T, which means that flux-pinning of GBs is still meaningful at the temperature. 
As temperature increases up to 20 K (Fig. \ref{fig4} (b)), diamagnetic property of the ACed specimen is larger than that of the WQed specimen in entire field except the low field. From the results, we can understand that  flux-penetrating promotion effect of  GBs was larger than  flux-pinning effect of  them at 20 K.

 The tendency is more pronounced at 25 K. Figure \ref{fig5} (a) shows that $\Delta$M of the ACed 5 wt.\% (Fe, Ti) doped MgB$_2$ are much larger than that of the WQed specimen over the entire field at 25 K. The decrease of diamagnetic property of the WQed specimen  is too great to compare with that of the ACed specimen. 
The tendency is much more pronounced at 30 K (Fig. \ref{fig5} (b)). The diamagnetic property of WQed 5 wt.\% (Fe, Ti) particle-doped MgB$_2$ specimens show even worse than that of ACed  pure MgB$_2$ specimen. The behavior proves the prediction that GBs did not play a role as  flux-pinning centers but did as flux-penetrating promotion pathway owing to increased coherence length at 30 K. On the other hand,  the maximum diamagnetic property of ACed doped specimen at 30 K is still not significantly different from the maximum  diamagnetic property of other temperatures. It is considered to be because leak-out depinning of the ACed doped specimen was small.

\subsection{Discussion}
\begin{figure}
\vspace{1cm}
\begin{center}
\includegraphics*[width=12cm]{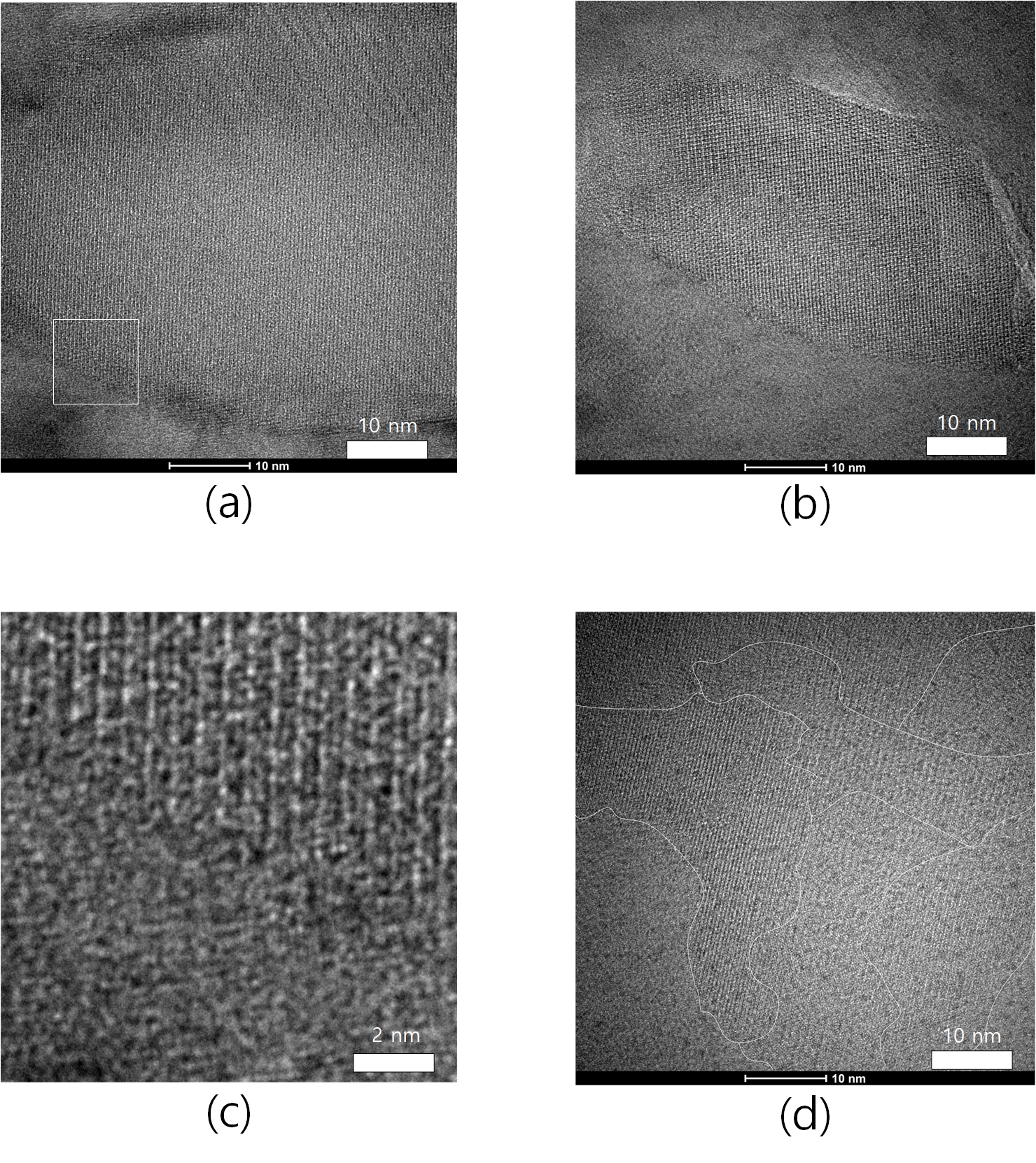}
\end{center}
\caption{TEM images of  air-cooled and water-quenched 5 wt.\% (Fe, Ti) particle-doped MgB$_2$ (a): Grain boundary of the air-cooled specimen. (b): Grain boundary of the water-quenched specimen. (c): Magnified image of (a). (d): Grain boundaries of the water-quenched specimen.} 
\label{fig6}
\end{figure}

In order to know the width of  GB and the amount of increased GBs by WQing, images of the specimens were taken by using transmission electron microscope (TEM).
TEM image of ACed 5 wt.\% (Fe, Ti) particle-doped MgB$_2$ is shown in Fig. \ref{fig6} (a). Inspecting GBs of MgB$_2$, they are rather clearly observed when  growth directions of two MgB$_2$ grains are definitely different, but GBs are not clearly observed when they are not. 
The image was taken for the area that GB is clearly shown. Considered by the image,  grain size of ACed specimen was over 100 nm. 

Figure \ref{fig6} (b) is TEM image of WQed 5 wt.\% (Fe, Ti) particle-doped MgB$_2$. A grain size of the WQed specimen was considerably reduced  when compared with that of ACed specimen, and the size of the grain is less than 50 nm.  
 Figure \ref{fig6} (c) is the magnified image of the square in  Fig. \ref{fig6} (a). It is observed that the width of GB is less than 1 nm and the distribution of atom at neighborhoods of the GB was a little distorted from regular array of MgB$_2$.

Therefore, a GB of MgB$_2$ cannot pin several quantum fluxes across the width of a GB because the width of a GB (1 nm) is narrower than the $\xi$ of MgB$_2$ (2.24 nm, 0 K). 
Another image of WQed 5 wt.\% (Fe, Ti) particle-doped MgB$_2$ are shown in  Fig. \ref{fig6} (d). We drew GB lines because GB lines of MgB$_2$ are not clearly shown because mismatch between grains are not significant, but it is clear that growth direction of grains is different in the image. It is certain that GB density  of WQed specimen considerably increased when compared with that of ACed specimen  as shown in Fig. \ref{fig6}.

Calculating increased  GB density by WQing, it is assumed that a grain of radius $r_o$ is divided to grains of radius r. Increased surface area by the divide is 
\begin{eqnarray}
\Delta A_{surface} = 4\pi r_o^2 \frac{4\pi (\frac{r_o}{n})^2}{4\pi r_o^2}\times n^3 = n \times 4\pi r_o^2
 \end{eqnarray} 
where n is the number of divide ($r_o$=$nr$).
However,  increased grain boundary area is $\frac{n}{2}$$\times 4\pi r_o^2$ because grains in a material are in contact with each other. Therefore,  GB density of the WQed specimen is 
\begin{eqnarray}
 A_{GB, WQ} = (1 + \frac{n-1}{2}) a_o
 \end{eqnarray}
where  $a_o$ is GB density of the ACed specimen.
Numerically, n is 2 when a grain size is reduced from 100 nm to 50 nm by WQing as shown in  Fig. \ref{fig6}. Thus, the  ratio of  GB density between the WQed specimen and the ACed specimen ($A_{GB, WQ}/A_{GB, AC}$) is $\frac{3}{2}$. 

 On the other hand,   increased GB density of the WQed specimen can be estimated by  using M-H curve.  Paying attention to 6 Tesla in Fig. \ref{fig2} (a).  $\Delta$M of  the WQed specimen is approximately two times compared with that of the ACed specimen. The flux-pinning effect of volume defects become infinitesimal at the field because volume defects already   have not only pinned the fluxes to their pinning limit but volume defects also are  same for the two specimen \cite{Lee4}. Thus, flux-pinning effects of GBs are dominant. 
 Considering  the amount of flux-pinning effects of GBs, n is 3 in WQed specimen by Eq. (3) because flux-pinning effect of the WQed specimen is two times compared with that of ACed specimen, which means that grains of the WQed specimen  are more refined than results of TEM.  We think that the result of M-H curves is more reliable than that of TEM because the images of TEM are focused at a area. 
 
\section{conclusion} 
We studied field dependence of magnetization for water-quenched 5 wt.\% (Fe, Ti) particles-doped MgB$_2$ specimen and compared with that of the air-cooled specimen for flux-pinning effects caused by increased  grain boundaries (GBs) by water-quenching. 
The diamagnetic properties of water-quenched specimen were rather lower than that of air-cooled specimen in less than 2.5 T at low temperatures (5 K and 10 K). It was because the increased GBs were served as a pathway leak-out depinning the fluxes pinned on the volume defects. On the other hand,  diamagnetic properties of water-quenched specimen were higher than that of air-cooled specimen in more than 2.5 T at the temperatures, which were because the increased GBs by water-quenching are served as pinning center. TEM revealed that the width of grain boundary in specimens is approximately 1 nm, which is not wide enough to pin several flux quanta across the width of GB. A representation revealed that a GB has  flux-pinning effect and a flux-penetrating promotion effect simultaneously. The increased coherence length of  the superconductor as temperature increases resulted in a significant decrease of the flux-pinning effect of water-quenched specimen compared with that of air-cooled specimen,  which was caused by increased GBs. On the other hand, degree of grain refinement by water-quenching for the specimen was evaluated by TEM and M-H curves, respectively. The result of M-H curves was more refined than that of TEM and the result of M-H curves is more reliable than that of TEM because the images of TEM are focused at a area. \\

 $\bf{Acknowledgment}$\\
The authors would like to thank Dr.  Byeong-Joo Kim of PNU for careful discussion.\\

\newpage

\begin{table}[!h]
\vspace{1cm}
\caption{ Ratios of  pinned fluxes at a planar defect along temperature, where $\xi_0$ is coherence length at 0 K and  $\xi_K$ is coherence length at a temperature. 
It is assumed that H$_{c2}$ of the superconductor is 65.4 T at 0 K,  T$_c$ is 37.5 K, and the width of grain boundary can pin a flux quantum.}
\begin{center}
\renewcommand{\tabcolsep}{8pt}
\renewcommand{\arraystretch}{0.3}
\begin{tabular}
{|c||c|c|c|c|c|c|c|c|}
\hline\hline
 & & & & & & & &\\
 Temperature &0 K &$5 K$& 10 K&15 K &20 K &25 K&30 K&35 K\\
  & & & & & & & &\\
  \hline\hline
   & & & & & & & &\\
 Coherence length (nm) &2.24  &$2.41 $& 2.62 &2.90 &3.29 &3.89&5.02&8.29\\
  & & & & & & & &\\
  Ratio of pinned fluxes    &1&$0.86$&0.73 &0.60 &0.47&0.33&0.20&0.07\\
on the grain boundary: $($$\frac{\xi_0}{\xi_K}$$)^2$ & & & & & & & &\\  & & & & & & & &\\
\hline
\end{tabular}
\end{center}
\end{table} 
\end{document}